\documentclass[twocolumn,showpacs,aps,prl,superscriptaddress]{revtex4}

\usepackage{graphicx}
\usepackage{dcolumn}
\usepackage{amsmath}
\usepackage{epsfig}
\usepackage{color}

\def\modeI{B^+ \rightarrow \pi^+ \pi^- \pi^+}
\def\modeII{B^+ \rightarrow K^+ \pi^- \pi^+}
\def\modeIII{B^+ \rightarrow K^+ K^- \pi^+}
\def\modeIV{B^+ \rightarrow K^+ K^- K^+}
\def\modeV{B^+ \rightarrow K^- \pi^+ \pi^+}
\def\modeVI{B^+ \rightarrow K^+ K^+ \pi^-}

\newcommand{\pvec}{{\bf p}}

\def\etal               {{\it et~al.,}}

\input babarsym

\newcommand{\BABARPubYear}    {03}
\newcommand{\BABARPubNumber}  {005}

\newcommand{\SLACPubNumber} {9703}

\def\figurebox#1#2#3{%
    \def\arg{#3}%
    \ifx\arg\empty
    {\hfill\vbox{\hsize#2\hrule\hbox to #2{\vrule\hfill\vbox to #1{\hsize#2\vfill}\vrule}\hrule}\hfill}%
    \else
    {\hfill\epsfbox{#3}\hfill}%
    \fi}

\begin{document}

\preprint{\babar-PUB-\BABARPubYear/\BABARPubNumber} 
\preprint{SLAC-PUB-\SLACPubNumber} 

\begin{flushleft}
\babar-PUB-\BABARPubYear/\BABARPubNumber\\
SLAC-PUB-\SLACPubNumber\\
\end{flushleft}

\title{
{\large \bf Measurements of the Branching Fractions and Bounds on the Charge Asymmetries of 
Charmless Three-Body Charged {\boldmath{$B$}} Decays} 
}

%
\author{B.~Aubert}
\author{R.~Barate}
\author{D.~Boutigny}
\author{J.-M.~Gaillard}
\author{A.~Hicheur}
\author{Y.~Karyotakis}
\author{J.~P.~Lees}
\author{P.~Robbe}
\author{V.~Tisserand}
\author{A.~Zghiche}
\affiliation{Laboratoire de Physique des Particules, F-74941 Annecy-le-Vieux, France }
% ...
\collaboration{The \babar\ Collaboration}
\noaffiliation

\date{\today}

\begin{abstract}
We present measurements of branching fractions and charge asymmetries
for charmless $B$-meson decays to three-body final states of charged
pions and kaons. The analysis uses 
$81.8~\invfb$ of data collected at the $\FourS$ 
resonance with the \babar\ detector at the SLAC \pep2\ asymmetric $B$ Factory.
We measure the branching fractions
$\BR(\modeI) = (10.9 \pm 3.3 \pm 1.6) \times 10^{-6}$,
$\BR(\modeII) = (59.1 \pm 3.8 \pm 3.2) \times 10^{-6}$, and
$\BR(\modeIV) = (29.6 \pm 2.1 \pm 1.6) \times 10^{-6}$, and provide 90\% C.L.
upper limits for other decays. We observe no charge asymmetries for these
modes.
\end{abstract}


\maketitle

The study of charmless hadronic $B$ decays can make important contributions 
to the understanding of \CP\ violation in the Standard Model,
as well as to models of hadronic decays.
Reference~\cite{ref:TheoryPaper} proposes that 
the interference between various charmless decays and the $\chi_{c0}$ 
resonance can be used to measure the Cabibbo-Kobayashi-Maskawa (CKM)
angle $\gamma$, while the decay $\modeI$ can be used to reduce the 
uncertainties in the measurement of the CKM angle $\alpha$~\cite{ref:alpha}.
We present branching fractions and charge asymmetries of
charged-$B$-meson decays to three-body final states of charged pions and 
kaons~\cite{ref:conjugates}, with no assumptions about intermediate
resonances and with charm contributions subtracted,
which allows us to set a tight bound on the charmless
contribution to the measurement of $\gamma$~\cite{ref:TheoryPaper}.
Upper limits and measurements of some of these
branching fractions have been obtained previously with
smaller statistics~\cite{ref:CLEOandBelle}.

The data used in this analysis were collected 
at the \pep2\ asymmetric \epem\ storage ring with the \babar\ detector,
described in detail elsewhere~\cite{ref:babar}. 
The on-resonance data sample consists of 88.8 million \BB\ pairs
collected at the \FourS\ resonance during the years 1999-2002.
We also use 9.6~\invfb\ of off-resonance data, collected 40~\mev\ below
the \FourS\ resonance, to characterize the backgrounds
from \epem\ annihilation into light \qqbar\ pairs. We assume that the
$\FourS$ decays equally to neutral and charged $B$ meson pairs.

Hadronic events are selected based on track multiplicity and event topology.
Backgrounds from non-hadronic events are reduced by requiring the ratio of 
Fox-Wolfram moments $H_2/H_0$~\cite{ref:FoxWolfram} to be less than 0.98.
Candidate $B$ decays are formed by combining three charged tracks, 
where each track is required to have at least
12 hits in the tracking chamber, a minimum transverse 
momentum of 100~\mevc, and to be consistent with originating from the beam-spot. 

Signal decays are identified using two kinematic variables:
1) the difference $\DeltaE$ between the center-of-mass (CM) energy of the
$B$ candidate and $\sqrt{s}/2$, where $\sqrt{s}$ is the total CM energy, and
2) the beam-energy substituted mass 
$\mes = \sqrt{(s/2 + \pvec_i \cdot \pvec_B)^2/E_i^2 - \pvec^2_B}$, 
where the $B$ momentum $\pvec_B$
and the four-momentum of the initial state ($E_i, \pvec_i$) are defined in the
laboratory frame.
The $\Delta E$ and $\mes$ distributions of signal events are Gaussian
with resolutions of $20~\mev$ and $2.7~\mevcc$, respectively.
The typical $\Delta E$ separation between modes that
differ by substituting a kaon for a pion in the final state is $45~\mev$, assuming
the pion mass hypothesis.

Charged pions and kaons are identified using energy loss
(\dedx) in the silicon detector and tracking chamber,
and, for tracks with momenta above 700~\mevc, 
the Cherenkov angle and number of photons measured by the Cherenkov detector.
The efficiency of selecting kaons is approximately 80\%,
which includes the geometrical acceptance,
while the probability of misidentifying pions as kaons is
below 5\%, up to a laboratory momentum of 4.0~\gevc. Pions are required to fail both
the kaon selection and an electron
selection algorithm based on information from \dedx, shower shapes in the 
calorimeter and the 
ratio of the shower energy and track momentum. 

We remove $B$ candidates when the invariant mass of the combination 
of any two of its daughter tracks (of opposite charge) 
is within $6\sigma$ of the mass of the 
$D^0$ meson or within $3\sigma$ of the mass of the
$J/\psi$, $\psi(2S)$ or $\chi_{c0}$ mesons~\cite{ref:pdg2002}.
Here, $\sigma$ is $10~\mevcc$ for $D^0$, $15~\mevcc$ 
for $J/\psi$ and $\psi(2S)$, and $18~\mevcc$ for $\chi_{c0}$.

To suppress background from light-quark and charm continuum
production, two event-shape variables are computed in the CM frame. 
The first is the cosine of the angle $\theta^*_T$ between 
the thrust axis of the selected $B$ candidate and the thrust axis of the rest 
of the event, i.e. all charged tracks and neutral particles
not assigned to the $B$ candidate.
For jet-like continuum backgrounds, $|$cos$\theta^*_T|$ is strongly peaked
towards unity, while it is essentially uniform for signal events.
For each signal mode we fix an upper limit on $|$cos$\theta^*_T|$, 
between 0.575 and 0.850. 
This rejects between 95\% and 85\% of the background, depending 
on the decay mode.

The second event-shape variable is a Fisher
discriminant~\cite{ref:Fisher}, which is formed from 
the summed scalar momenta of all charged and 
neutral particles from the rest 
of the event within nine $10^\circ$-wide nested cones coaxial with the 
thrust axis of the $B$ candidate. The coefficients of the Fisher discriminant
are chosen to maximize the separation between signal and continuum 
background events, and are calculated for each signal mode separately using
Monte Carlo simulated signal and continuum events. A further 50\% to
75\% of the remaining background is rejected, depending on the decay mode, 
by applying selection requirements on this variable.

\begin{figure}[!htb]
\begin{center}
\mbox{\epsfig{file=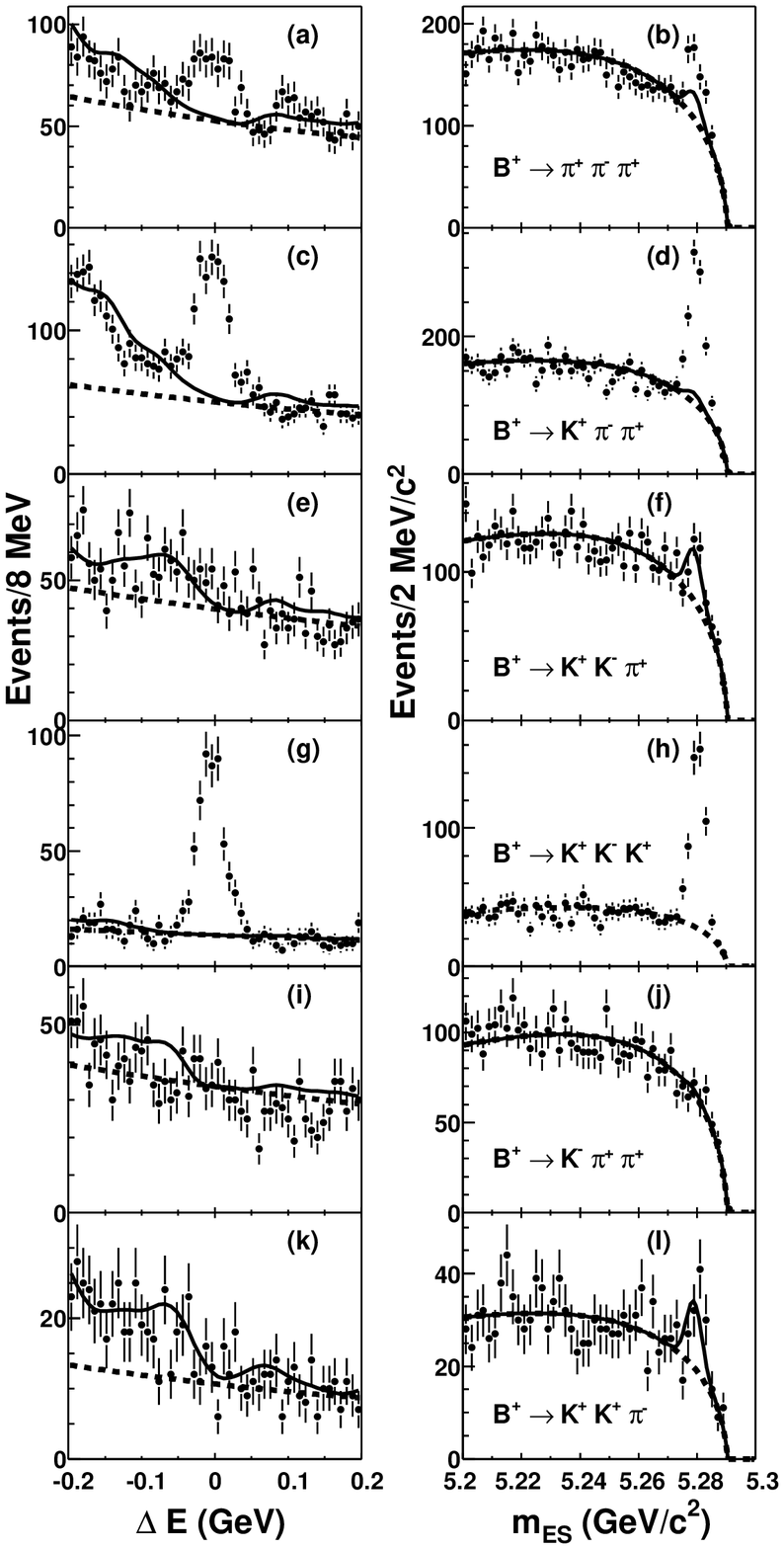, bb=15 100 350 760, clip=, width=\linewidth, angle=0}}
\end{center}
\vspace*{-0.5cm}
\caption
{Projections of $\DeltaE$ and $\mes$ for $\modeI$ (a and b),
$\modeII$ (c and d), $\modeIII$ (e and f), $\modeIV$ (g and h),
$\modeV$ (i and j) and $\modeVI$ (k and l) in the signal region. The signal region
requirement was made on the orthogonal variable in each case. The
dashed curves show the continuum background, while the solid lines include
the \BB\ background.
\label{PRLdEandmESPlots}}
\end{figure}

$B$ decay candidates passing the above selection criteria are required to lie in
a signal region defined as follows: $|\mes - m_{B}| < 8~\mevcc$ and 
$|\DeltaE - \left<\DeltaE\right>| < 60~\mev$, 
where $\left<\DeltaE\right> = 7~\mev$ is the mean
value of $\DeltaE$ measured from on-resonance data for the calibration sample
$B^- \ra D^0 \pi^-, D^0 \ra K^- \pi^+$, and $m_{B}$ 
is the nominal mass of the charged $B$ meson~\cite{ref:pdg2002}.
Figure~\ref{PRLdEandmESPlots} shows the projections of the on-resonance data in
the signal region onto the $\DeltaE$ and $\mes$ axes.
Each plot shows the expected levels of continuum and \BB\ background, where the latter 
is parameterized from Monte Carlo samples.

The residual continuum background level is
estimated from the observed number of events in the grand sideband (GSB) region, 
defined to be $5.21 < \mes < 5.25~\gevcc$ and 
$|\DeltaE - \left<\DeltaE\right>| < 100~\mev$,
and extrapolating into the signal region.
The shape of the $\mes$ distribution of the background 
is parameterized according to the 
phenomenologically motivated ARGUS function~\cite{ref:Argus},
and is measured using off-resonance data and 
the upper sideband in the $\DeltaE$ variable 
in on-resonance data ($0.10 < \DeltaE < 0.25\gev$). 
A quadratic function is used to parameterize the \DeltaE distribution 
of the background.
The ratio of the integrals over the signal and GSB regions of the product
of the $\DeltaE$ and $\mes$ shape functions, $R$, gives the ratio
of the number of background events in the two areas.

The branching fraction for each channel is measured over the whole Dalitz plot,
which is divided into $28\times28$ cells of equal area $(1~\gev^2/c^4)^2$ 
to enable us to find the 
selection efficiency as a function of position in the Dalitz plot. 
\begin{table*}[!htb]
\caption{
\label{tab:results}
Branching fraction results for on-resonance data. The quantities
and their uncertainties are explained in the text.}
\vspace*{-0.5cm}
\begin{center}
\begin{tabular}{|l|cccccc|}
\hline
\hline
Signal Mode & $\pi^{\pm} \pi^{\mp} \pi^{\pm}$ & $K^{\pm} \pi^{\mp} \pi^{\pm}$ 
& $K^{\pm} K^{\mp} \pi^{\pm}$ & $K^{\pm} K^{\mp} K^{\pm}$ & $K^{\mp} \pi^{\pm} \pi^{\pm}$ & $K^{\pm} K^{\pm} \pi^{\mp}$ \cr
\hline
$\sum_i N_{1i}$ & 1029 & 1502 & 733 & 646 & 494 & 209 \cr
$\sum_i N_{2i}$ & 5577 & 5209 & 4012 & 1308 & 3268 & 1025 \cr
$\left<\epsilon\right>$(\%) & $12.7 \pm 0.5$ & $12.8 \pm 1.4$ & 
$13.9 \pm 0.9$ & $14.9 \pm 0.9$ & $18.5 \pm 0.9$ & $15.3 \pm 0.7$ \cr
$R$ & $0.144 \pm 0.003$ & $0.146 \pm 0.003$ & $0.150 \pm 0.003$ & $0.158 \pm 0.006$ 
& $0.155 \pm 0.003$ & $0.157 \pm 0.006$ \cr
\hline
A) $\sum_i N_{1i}/\epsilon_i$ 
& $7597 \pm 275$ & $11056 \pm 327$
& $5071 \pm 216$ & $4011 \pm 182$ & $2670 \pm 120$ & $1366 \pm 94$\cr
B) $\sum_i RN_{2i}/\epsilon_i$ 
& $5938 \pm 94 \pm 117$ & $5604 \pm 89 \pm 111$
& $4041 \pm 72 \pm 80$ & $1381 \pm 46 \pm 55$ & $2738 \pm 48 \pm 53$ & $1052 \pm 33 \pm 40$ \cr
C) $\sum_i N_x p_i/\epsilon _i$ 
& $474 \pm 33 \pm 40$ & $22 \pm 1 \pm 30$ & $671 \pm 15 \pm 59$ & --- & --- & $344 \pm 31$ \cr

D) $n_x$ & --- & $-189 \pm 34$ 
& $110 \pm 128$ & --- & --- & $53 \pm 5$ \cr
E) $D^0$ Bkgnd & $216 \pm 24$ & $268 \pm 28$& $47 \pm 6$ & --- & $33 \pm 5$& $31 \pm 5$\cr
F) $\eta' K$ Bkgnd & --- & $106 \pm 30$ & --- & --- & --- & --- \cr
G) Signal Yield & $970 \pm 291 \pm 130$ & $5246 \pm 339 \pm 127$
& $202 \pm 227 \pm 163$ & $2630 \pm 188 \pm 55$ & $-101 \pm 129 \pm 53$ & $-114 \pm 100 \pm 51$\cr
& $\pm 22 \pm 50$& $\pm 39 \pm 247$ & $\pm 16 \pm 9$ & $\pm 12 \pm 124$ & $\pm 0 \pm 5$ & $\pm 0 \pm 5$ \cr
\hline
$\BR$ ($\times 10^{-6}$) 
& $10.9 \pm 3.3 \pm 1.6$ & $59.1 \pm 3.8 \pm 3.2$
& $2.3 \pm 2.6 \pm 1.8$ & $29.6 \pm 2.1 \pm 1.6$ & $-1.1 \pm 1.5 \pm 0.6$ & $-1.3 \pm 1.1 \pm 0.6$ \cr
Significance ($\sigma$) & 5.7 & $> 6$ & 1.1 & $> 6$ & --- & --- \cr
90\% C.L.& --- & --- & $<6.3$ & --- & $<1.8$ & $<1.3$ \cr
\hline
\hline
\end{tabular}

\end{center}
\end{table*}
Taking $\epsilon_i$ to be the efficiency of reconstructing the signal in the
$i^{th}$ bin of the Dalitz plot, determined from Monte Carlo simulated events, the 
branching fraction for each signal mode is given by:
\begin{equation}
{\cal{B}} = \frac{1} {N_{\BB}} \left(\sum_i \frac{(N_{1i} - R N_{2i} - N_xp_i)} {\epsilon_i} - n_x - n_{b} \right),
\label{eqn:firstBR}
\end{equation}
where $N_{1i}$ and $N_{2i}$ are the number of events observed in the 
signal and GSB regions, respectively, while $N_x p_i$,
$n_x$ and $n_b$ are background contributions that are defined below.
$N_{\BB}$ is the total number of $\BB$ pairs in the data sample.
No significant differences were 
found for the value of $R$ (defined earlier)
in different regions of the Dalitz plot, so an average value is used for all bins.

The probability of a kaon being misidentified as a pion
is 20\%. 
This means there is significant
cross-feed into the signal region from the decay mode that has one more
kaon, which is subtracted for each bin, $i$. This is represented
by the $N_xp_i/\epsilon_i$ term in Eq.~(\ref{eqn:firstBR}), where
$N_x$ is the total number of events that is the source of the cross-feed, and
$p_i$ is the probability for the cross-feed events to pass the 
selection criteria for the $i^{th}$ bin, which is estimated from Monte Carlo samples.
The $\modeIV$ mode has $N_x = 0$, since it has no cross-feed backgrounds. For the
other decays, $N_x$ is obtained by multiplying $N_{\BB}$ by the 
branching fraction of the signal mode that
has a kaon substituting a pion in the final state.
There is also second-order cross-feed where either two kaons
are misidentified as pions (probability of 4\%), or one of the pions is 
misidentified as a kaon (probability of 2\%). 
This is represented by the $n_x$ term in Eq.~(\ref{eqn:firstBR}).

Finally, the $n_{b}$ term represents the
small number of other \BB\ backgrounds that are subtracted.
For all signal modes except $\modeIV$, $n_b$ is obtained from
the number of $D^0$ and $\bar{D}^0$ candidates whose invariant mass is beyond
the $6\sigma$ range. For $\modeII$, there is also a contribution from 
$B^{\pm} \ra \eta' (\ra \rho^0 \gamma) K^{\pm}$ decays, which is estimated 
from the selection efficiency from Monte Carlo simulated decays,
and the branching fraction quoted in Ref.~\cite{ref:pdg2002}. 
By using a mixture of Monte-Carlo-simulated charm and charmless decays, we
found that there were no other significant \BB\ backgrounds.

We do not divide the Dalitz plot into cells for the 
Standard-Model-suppressed modes $\modeV$ and $\modeVI$, and instead
use the average values of the signal efficiency and cross-feed terms.

The branching fraction results are summarized in Table~\ref{tab:results}, where
the first four rows show the total number of events in the signal
and GSB regions, the average signal efficiencies
$\left<\epsilon\right>$, and the values of $R$ for each mode. The absolute
efficiency variation across the Dalitz plot typically varies between $\pm$2\% 
and $\pm$5\% from $\left<\epsilon\right>$.

Rows A and B represent the total number of events and the amount of 
continuum background in the signal region, corrected for efficiency.
The uncertainties for row A come from
the statistical errors in $N_{1i}$, while the uncertainties
for row B correspond to the
statistical errors in $N_{2i}$, and the systematic errors
from $R$, which arise from the limited statistics in the
sideband region and off-resonance data.

Row C shows the expected background from cross-feed events.
The first and second uncertainties
of these quantities represent the systematic errors in
$p_i$ and $N_x$, respectively, except for the channel $\modeVI$, 
where the uncertainty represents the average of the 
$p_i$ and $N_x$ contributions.
The second-order cross-feed terms $n_x$ are shown in row D.
Note that the $n_x$ term for $\modeII$ is negative, which
corrects for the $\modeIV$ cross-feed into $\modeIII$, which in turn contributes
to the cross-feed background for $\modeII$.

Rows E and F show the expected backgrounds from 
$D^0$ and $\eta'K$ decays, which include the
uncertainties from the selection efficiencies and the branching fractions of
the background decays~\cite{ref:pdg2002}.
The sum of these two rows gives the value of $n_{b}$ in Eq.~(\ref{eqn:firstBR}).

Row G shows the signal yield, obtained by subtracting rows B to F from
row A. The first uncertainty is the combination of the statistical
errors of the number of events in the signal and GSB regions.
The second uncertainty corresponds to the sum in quadrature
of all the systematic errors from rows B to F. The third error
is from the bin-by-bin uncertainty of the selection efficiency.
This is zero for $\modeV$ and $\modeVI$, where we only 
use the average efficiencies. The last uncertainty originates from global systematic 
errors in the signal efficiencies due to charged-particle
tracking (0.8\% per track), event-shape variable selections 
(1.0 to 2.5\%) and from particle identification
(1.4\% and 1.0\% per pion and kaon track, respectively).

The next row in Table~\ref{tab:results} shows the branching fraction results, 
where the first uncertainties are from the statistical errors
in the number of events, while the second uncertainties
are the sum in quadrature of all systematic errors mentioned above.

The significance of each branching fraction result, under the null hypothesis,
is defined as the ratio of the signal yield to the total 
(statistical and systematic)
uncertainty of the background in the signal region.
We observe significant signals for the modes
$\modeI$, $\modeII$ and $\modeIV$, and provide 90\% C.L. upper limits
for the other channels, using the formalism in Ref.~\cite{ref:FeldmanCousins}.
The branching fraction of the control sample
$B^- \ra D^0 \pi^-, D^0 \ra K^- \pi^+$, which has the same final state
as $\modeII$, is measured to be
$(190 \pm 3 \pm 10) \times 10^{-6}$, which agrees with the 
average of published measurements
$(201 \pm 20) \times 10^{-6}$~\cite{ref:pdg2002}.

We have also measured the charge asymmetries for the modes with observed
signals using a method similar to that used for the branching fraction
measurements. The charge asymmetries are defined as
${\cal{A}} = (N^- - N^+)/(N^- + N^+)$, where $N^-$ ($N^+$) is the
signal yield for negatively (positively) charged $B$ candidates, as defined
by row G in Table~\ref{tab:results}.
The normalisation factor $N_{\BB}$ cancels
out in the asymmetry ratio, while the cross-feed
and $\BB$ background contributions cancel in the asymmetry 
numerator. The measured charge asymmetries are
${\cal{A}}(\modeI) = -0.39 \pm 0.33 \pm 0.12$, 
${\cal{A}}(\modeII) = 0.01 \pm 0.07 \pm 0.03$ and 
${\cal{A}}(\modeIV) = 0.02 \pm 0.07 \pm 0.03$,
where the first uncertainties are statistical and the second 
are systematic, which include the 
charge bias of the tracking and particle identification selection
requirements (1\%).

In summary, we have measured the branching fraction of $\modeI$
for the first time, which is smaller than that assumed
in Ref.~\cite{ref:TheoryPaper},
and we have also observed the channels $\modeII$ and $\modeIV$. 
We observed no charge asymmetries in these decays. 

We are grateful for the excellent luminosity and machine conditions
provided by our \pep2\ colleagues, 
and for the substantial dedicated effort from
the computing organizations that support \babar.
The collaborating institutions wish to thank 
SLAC for its support and kind hospitality. 
This work is supported by
DOE
and NSF (USA),
NSERC (Canada),
IHEP (China),
CEA and
CNRS-IN2P3
(France),
BMBF and DFG
(Germany),
INFN (Italy),
FOM (The Netherlands),
NFR (Norway),
MIST (Russia), and
PPARC (United Kingdom). 
Individuals have received support from the 
A.~P.~Sloan Foundation, 
Research Corporation,
and Alexander von Humboldt Foundation.

\end{document}